# Synthesis and conformational analysis of new derivatives of 7−Chloro-1, 3-Dihydro-5-Phenyl-2H-1, 4-Benzodiazepine-2-One

**Gholamhassan Imanzadeh, Zahra ArastehFard*, Yaser Sadra**

**Department of Sciences, University of Mohaghegh Ardabili, Ardabil, (IRAN)**
**E-mail: z.arastehfard@msn.com**

**ABSTRACT**

1,4-benzodiazepine-2-ones and their derivatives are prominent structures in medicinal chemistry. These biomolecules have wide biological activities and posses therapeutic applications. In this works, we introduce new derivatives of 1,4-benzodiazepine-2-ones which are synthesized using michael addition reaction of 7-chloro- 1,3-dihydro-5-phenyl-2H-1,4-benzodiazepine-2-ones with fumaric esters that matches with green chemistry protocols. The structures of all products are confirmed by FT-IR, $^1$H-NMR, $^{13}$C-NMR and MASS spectroscopy. Since the stereochemistry of 1,4-benzodiazepine-2-ones is important, we study the most stable conformer of one of the products as a model for conformational analysis by hyper chem soft ware and semi empirical $AM_1$ program. Also, using the $^1$H-NMR spectrum, we investigate the produced diastereomers of one of products as a model.

**INTRODUCTION**

1,4-benzodiazepine-2-ones are important biomolecules with a wide area of biological activities and have therapeutic applications[1]. The activities of 1,4-benzodiazepine-2-ones be known in Central Nervous System (CNS)[2]. These activities included anxiolytic, anticonvulsant, sedative, hypnotic and muscle relaxants[3]. Also, their activities be confirmed as agents of anti-HBV, anti-HIV, anti ischemic, anticancer , antiarrhythmic of heartbeat and so on[4-8]. Accordingly, various derivatives of these biomolecules were synthesized[9]. In this work, we introduce new derivatives of these biomolecules which are synthesized using michael addition reaction of 7-chloro-1,3-dihydro-5-phenyl-2H-1,4-benzodiazepine2-ones with α,β-unsaturated esters (the symmetrical fumaric esters). The michael addition reaction is one of the most useful methods for forming carbon-carbon bond and has wide synthetic applications. From the view point of pharmaceutical and biological, the michael addition reactions on the α,β-unsaturated esters are important[10], we synthesize the new derivatives in the absence of solvent, under conventional heating conditions. These reactions carried out in the presence of 1,4-diazabicyclo [2.2.2]octane (DABCO). This is a non-toxic base and a good solid catalyst. Tetra Butyl Ammonium Bromide (TBAB) is quaternary ammonium salt which is used as a green media. Conformational analysis is the study of the energies and structures of conformations of organic molecules and their chemical and physical properties. Organic chemists use conformational analysis to understand the behavior of molecules in chemical reactions[11]. Here, because the importance of stereochemistry of 1,4-benzodiazepine -2-ones, we study the most stable conformer of one of the products as a model for conformational analysis by hyper chem and semi empirical $AM_1$ programs.

## EXPERIMENTAL

All chemicals were obtained from Merck or Fluka Chemical Companies. Some α,β-unsaturated esters were prepared from the corresponding fumaric acids by the reported method[12] and their structures were confirmed by FT-IR and $^1$H-NMR spectra. The progress of the reactions was followed by TLC using silica gel SILG/UV 254 plates. The FT-IR spectrum was recorded using a Perkin Elmer FT-IR spectrum RX-I spectrometer . The $^1$H-NMR (500 MHz) and $^{13}$C-NMR (125 MHz) spectra were recorded on Bruker 500 MHz Ultra Shield. The chemical shifts are given in ppm (δ) relative to internal TMS and coupling constants J are reported in Hz. The mass spectra were recorded with a Finnigan-MAT 8430 mass spectrometer operating at an ionization potential of 70 eV. The melting points were determined in open capillaries with a stuart melting point apparatus and are uncorrected. The column chromatography was performed on Merck silica gel 60 (230-240 mesh).

- **Preparation methods of 7-chloro-1,3-dihydro-5-phenyl-2H-1,4-benzodiazepine-2-one**

  As michael donor of reaction, we prepared 7-chloro-1,3-dihydro-5-phenyl-2H-1,4-benzodiazepine-2-one with two methods. In first method, we warmed 2-amino-5-chloro-benzophenon (0.5 g, 2.16 mmol) with excess amount from glycine ester hydrochloride (98 %) (0.6 g, 4.32 mmol) in dry pyridine. We removed some of the water and alcohol of the reaction in order to obtain the desired 1,4-benzodiazepine-2-one. We obtained the pure crystals of 7-chloro- 1,3-dihydro-5-phenyl-2H-1,4-benzodiazepine-2-one using purified by column chromatography[13]. Second method include three steps. In first step, from reaction of 2-amino-5-chloro-benzophenon (0.6 g, 2.3 mmol) with boromo acetyl bromide (0.26ml, 3mmol), we obtained α-boromoacetamid intermediate as white powder . Using TLC, we observed that almost all the initial material be converted to α-boromo acetamid intermediate. In second step, we used the aqueous solution of ammonia 25% ($\frac{w}{w}$) (18ml) and methanol (10ml) for ammunition of α-boromoacetamid intermediate. The reaction continued overnight until be obtained compound of amino acetamido. In third step, cyclization be performed under acidic conditions by reflux with acetic acid 10% ($\frac{v}{v}$) in tert-Butanol (15ml). The pure crystals of 1,4-benzodiazepine-2-one be obtained using purified by column chromatography[14]. Melting point of the obtained products were 214°C. Also, structure of the obtained products were confirmed by FT-IR and $^1$H-NMR.

- **Reaction of 1,4-benzodiazepine-2-one with symmetrical fumaric esters in solvent-free**

  We added α,β-unsaturated ester (1.3 mmol) to a well ground mixture of 7-chloro-1,3-dihydro-5-phenyl-2H-1,4-benzodiazepine-2-one (0.3 g,1 mmol), TBAB (0.54 g, 1.7 mmol) and DABCO (0.19 g, 1.7 mmol) in a 10mL balloon with a special glass door . The resulting mixture was heated in a heating oven at 70°- 90° C for 20 hours. Next, the reaction mixture was cooled in the room temperature and we dissolved in chloroform (10 ml). We filtered and washed the filtrate with water (2 × 10 ml) and then dried with $CaCl_2$. The solvent was evaporated and the crude product was purified by column chromatography on silica gel with EtOAc/n-hexane (1/10). The structures of all products are confirmed by FT-IR, $^{1}$H-NMR, $^{13}$C-NMR and MASS spectroscopy.

- **Physical and spectroscopic data of isolated products**

(1) FT-IR νmax(KBr) / $cm^{-1}$ : 3179.3, 3042.7, 2961,6, 1682.8, 1607.3, 1479.9.
$^{1}$H-NMR ($CDCl_3$) δppm: 4.34 (s, 2H), 7.148-7.555 (m, 8H), 9.40 (s, 1H).

(2)FT-IR νmax(neat) / $cm^{-1}$ : 3254.6, 2982.3, 2937.9, 1710.9, 1610.4, 1418.3.
$^{1}$H-NMR ($CDCl_3$) δppm: 1.18-1.32 (m, 12H), 3.09(d, 1H, *J* = 7 Hz), 3.85 (q, 1H, *J* = 7 Hz), 4.16 (d, 1H, *J* = 7 Hz), 5.07-5.11 (m, 2H), 6.80-7.50 (m, 8H), 9.89 (s, 1H).
$^{13}$C-NMR ($CDCl_3$) δppm: 22.2, 33.9, 43.9, 64.5, 68.7, 123.3, 128.7, 128.8,128.9, 130.2, 130.2, 130.8, 131.1, 132.4, 134.3, 137.3, 138.7, 168.4, 171.2, 172.7, 172.8.
MS m/z: 470 (7.00, $M^+$), 399 (21.50), 167 (35.80), 149 (100.00), 99 (50.00), 84 (78.60), 57 (50.00), 43 (100.00).

(3) FT-IR νmax(neat) / $cm^{-1}$: 3259.0, 3077.0, 2959.0, 1734.0, 1689.0, 1608.0, 1480.0.
$^{1}$H-NMR ($CDCl_3$) δppm: 0.86-0.95 (m, 6H), 1.24-1.40 (m, 4H), 1.54-1.63 (m, 4H), 3.12-3.17 (m,2H), 3.96-3.99 (m, 1H), 4.06-4.18 (m, 5H), 7.30-7.53 (m, 8H), 9.65 (s, 1H).
$^{13}$C-NMR ($CDCl_3$) δppm: 13.7, 19.0, 30.6, 33.2, 42.1, 62.8, 64.6, 64.8, 123,0, 128.3, 128.4, 128.5, 128.9, 129.1, 129.8, 130.5, 130.9, 132.1, 137.08, 138.3, 170.8,172.0, 173.9.
MS m/z: 499 (100.00, $M^+$+1), 425 (27.30), 322 (45.50), 241 (36.40), 77 (9.00), 57 (54.40), 41(81.80).

(4)FT-IR νmax(neat) ) / $cm^{-1}$: 3227.9, 3003.5, 3959.7, 1713.2, 1610.5, 481.3.
$^{1}$H-NMR ($CDCl_3$) δppm: 0.86-0.92 (m,6H), 1.27-1.37 (m, 8H), 1.58-1.67 (m, 4H), 3.13 (d, 2H, *J* = 7Hz),3.85-3.87 (m, 1H), 4.08-4.18 (m, 5H), 7.11-7.52 (m, 8H), 9.17 (s, 1H).
$^{13}$C-NMR ($CDCl_3$) δppm: 13.9, 22.3, 28.0, 28.1, 28.2, 28.3, 33.1, 43.4, 63.9, 64.81, 65.1, 122.7, 128.3, 128.6, 129.1, 129.9, 130.6, 130.8, 132.0, 136.7, 138.3,167.9, 170.5, 172.9, 172.9.
MS m/z: 526 (36.40, $M^+$), 411 (45.50), 322 (81.90), 295 (36.50), 241 (54.60), 83 (36.50), 71 (63.60), 57 (90.90), 43 (100.00).

(5)FT-IR νmax(neat) ) /$cm^{-1}$: 3227.9, 3003.5, 3959.7, 1713.2,1610,5, 1481.3.
$^{1}$H-NMR ($CDCl_3$) δppm: 0.84-0.92 (m, 6H), 1.23-1.34 (m, 12H), 1.57-1.66 (m, 4H), 3.11 (d, 2H, *J* = 7 Hz), 3.86 (q, 1H, *J* = 7 Hz), 4.01-4.17 (m, 5H), 7.13-7.50 (m, 8H) , 9.68 (s, 1H).
$^{13}$C-NMR ($CDCl_3$) δppm: 12.9, 21.5, 24.6, 27.5, 30.4, 32.2, 42.4, 62.9, 63.8, 64.4, 121.8, 126.8, 127.3, 127.4, 128.0, 128.8, 129.5, 129.7, 131.0, 135.8, 137.3, 140.3, 166.9, 169.7, 171.9, 172.0.

MS m/z: 554 (86.70, $M^+$), 424 (60.00), 322 (80.00), 270 (66.70), 227 (60.00), 43 (100.00).

## RESULT AND DISCUSSION

1,4-benzodiazepine-2-ones are to the form of a seven-member boat that substitutions on carbon $C_3$ (number of 3) are located in axial and equatorial situations[15]. With generation of a chiral center in $C_3$, conformational equilibrium of the molecules shifts toward conformer with major substitution in equatorial situation[16, 17]. In the $^1$H-NMR spectra, signal related to hydrogens of the $C_3$ be appeared in δ=4.43 ppm and as singlet. Inversion barrier of ring of benzodiazepine is 12.3 ($\frac{kcal}{mol}$). This value does not suffice for dissociation of methylene protons in axial and equatorial situations of the ring in NMR time scale and room temperature. Because, conformers (a) and (b) be converted together . Consequently, racemization occurs rapidly (see Figure 1)[18].

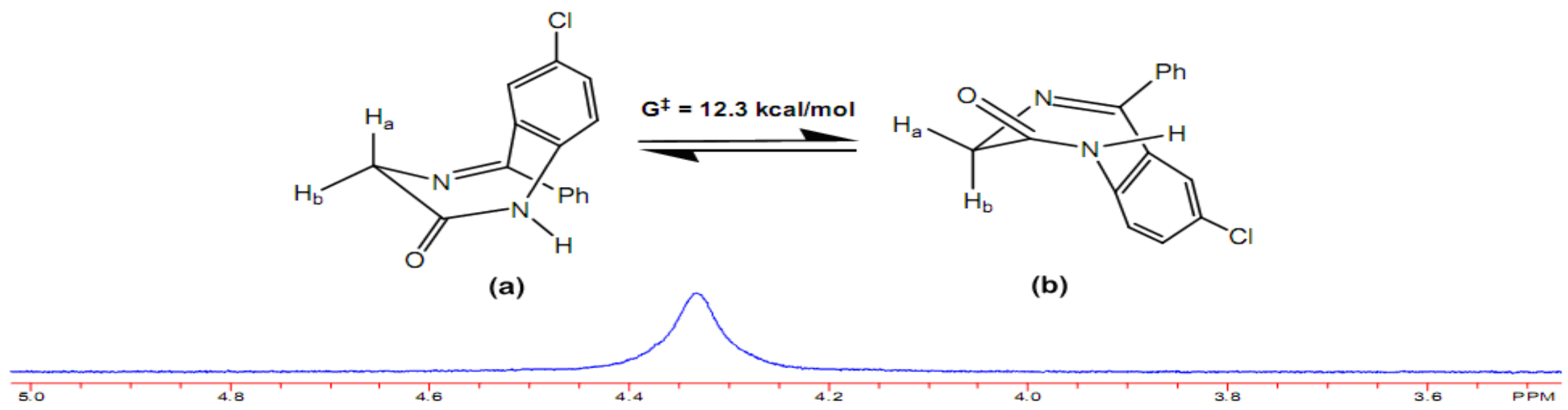

**Figure 1: $^1$H-NMR studies on 7-chloro- 1,3-dihydro-5-phenyl-2H-1,4-benzodiazepine-2-one.**

In this work, we investigated the addition reaction of 7-chloro-1,3-dihydro-5-phenyl-2H-1,4-benzodiazepine-2-one with symmetrical fumaric esters. Mild Organic base used,is DABCO that be used as a catalyst. This reaction is done in solvent-free, under thermal conditions and in presence TBAB as highly efficient and green alkylating agent. The reaction does not perform in the absence of TBAB. Interestingly, $^1$H-NMR spectra of all products shows that firstly, product of michael addition reaction is C-alkylation and is not N-alkylation. Secondly, alkyl groups occupy equatorial situation instead of axial situation. We obtained products to good yields (see Figure 2). After the study of products spectra, the results are presented in the Figure 3 investigations of various show that optimum temperature of reaction is in range 70°- 90°C and best time of performance of the reaction in thermal conditions is 20 hr.

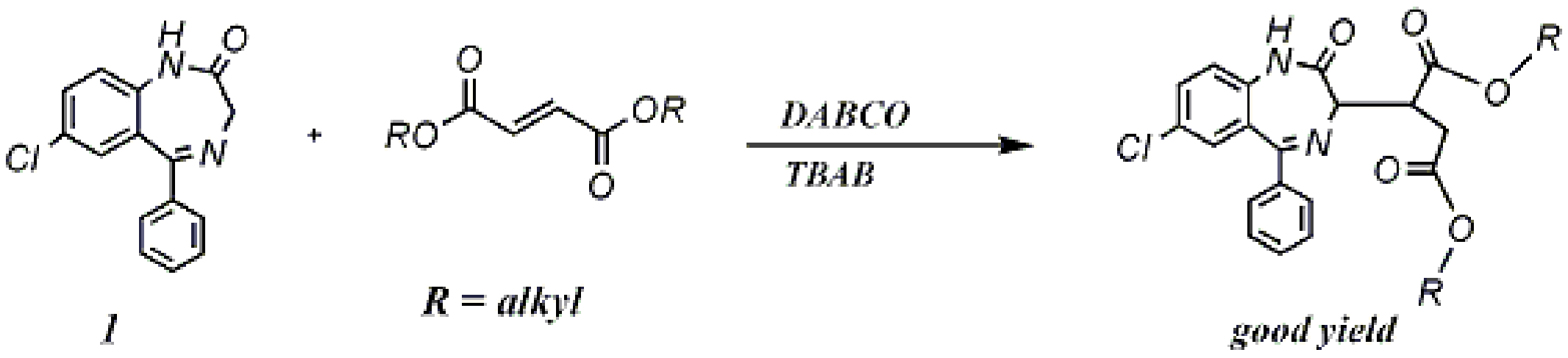

**Figure 2: Michael addition reaction of 7-chloro-1,3-dihydro-5-phenyl-2H-1,4-benzodiazepine-2-ones (1) with fumaric esters.**

| Entry | Ester | Product |
|---|---|---|
| 1 | 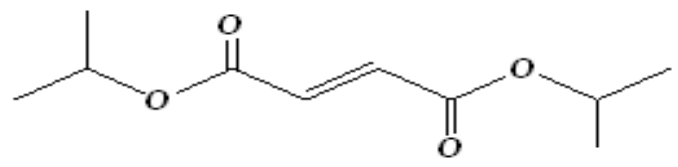 | 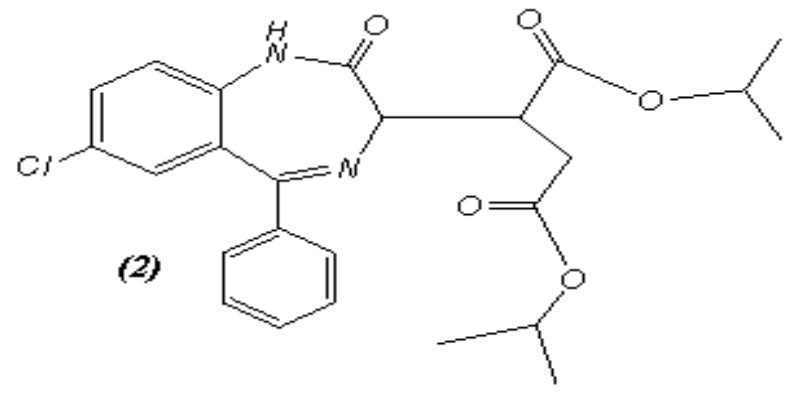 |
| 2 | 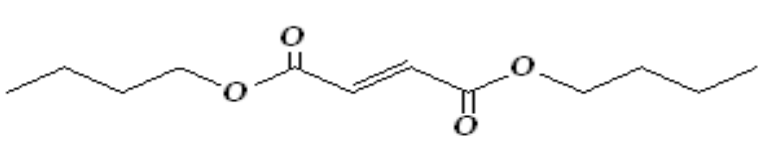 | 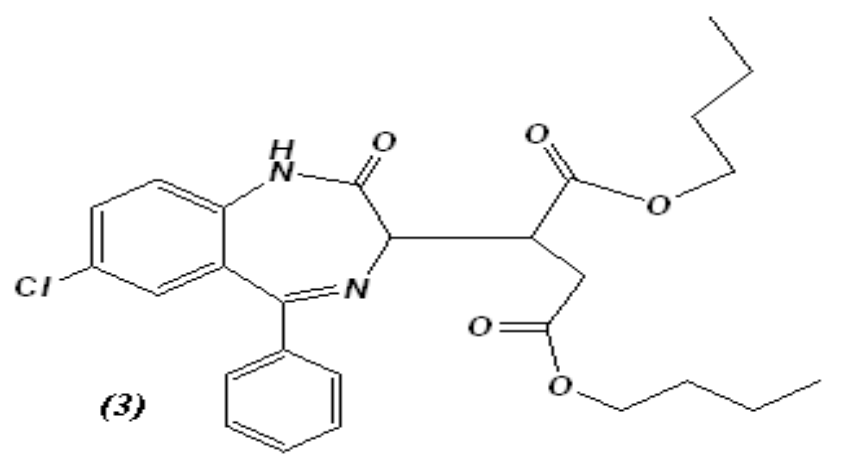 |
| 3 | 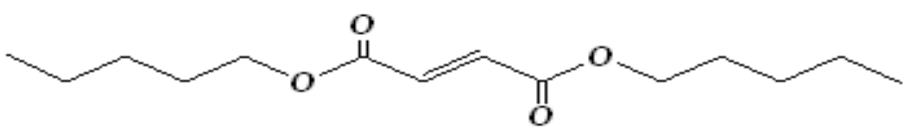 | 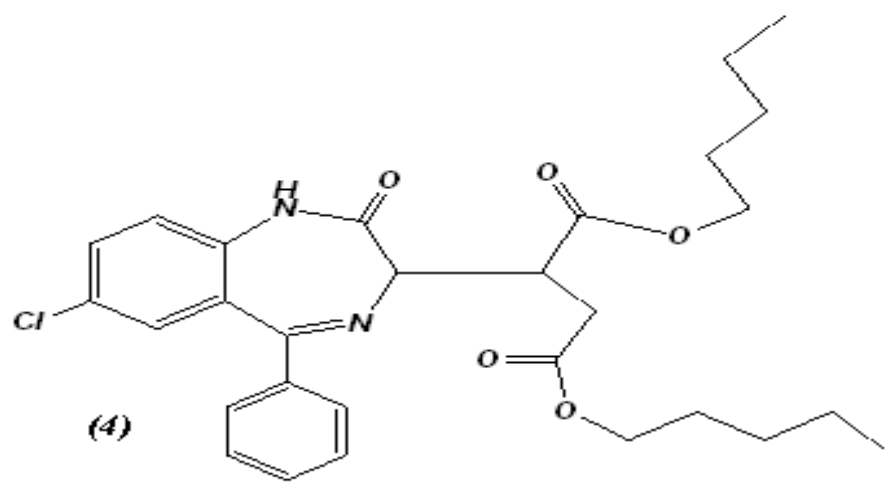 |
| 4 | 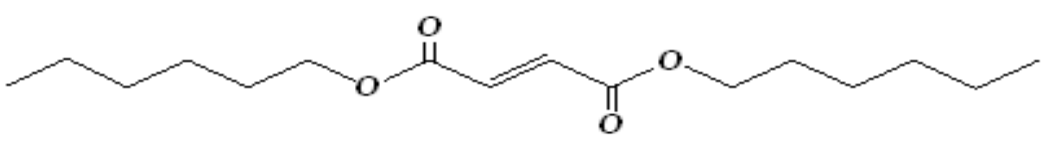 | 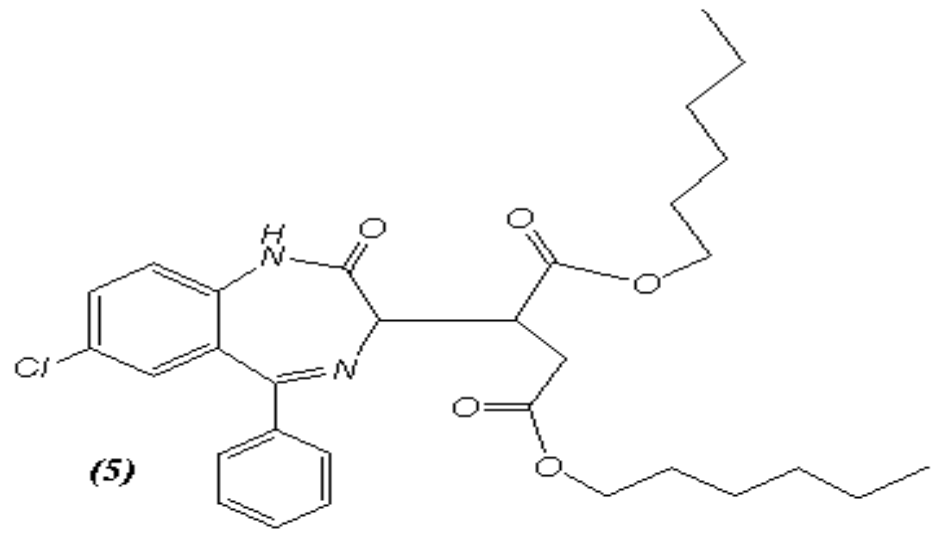 |

**Figure 3: Michael addition reactions of 7-chloro-1,3-dihydro-5-phenyl-2H-1,4-benzodiazepine-2-one with fumaric esters in presence of DABCO and TBAB.**

- **Use of semi empirical $AM_1$ program in investigation proposed hypothetical model**

Here, we study conformational analysis of one of the products by hyper chem soft ware and semi empirical $AM_1$ program. The study shows in C-alkylated form, obtained Products are as boat conformer with alkyl substitution in equatorial situation and are more stable than axial form. Despite hydrogen attached to amid nitrogen is more acidic than hydrogens of carbon $C_3$, alkylation occurs with fumaric ester in $C_3$. In order to investigate this subject, we study optimization energy of N-alkylated hypothetical molecule using semi empirical $AM_1$ program. The results show that in N-alkylated form, Because the steric crowding between hydrogen of phenyl group adjacent with nitrogen of amid (Here, their distance is 2.3 A°) is obtained the more unstable product. while in C-alkylation does not exist the steric crowding. As a model, for the additional reaction of 7-chloro-1,3-dihydro-5-phenyl-2H-1,4-benzodiazepine-2-one with hexyl fumarat, the energy difference between products of C-alkylation and N-alkylation is 17.6 ($\frac{kcal}{mol}$). Therefore, we propose that in 7-chloro-1,3-dihydro-5-phenyl-2H-1,4-benzodiazepine-2-one molecule, boat conformation is as only the stable conformer of molecule. In the conformer , atom $N_1$ is located so that the N-alkylation reaction with the symmetrical fumaric esters make impossible. The results is confirmed by the spectrum of $^1$H-NMR of products. Also, the results show the stable isomer is that its larger substitution located in equatorial situation (see Figure 4 and Figure 5).

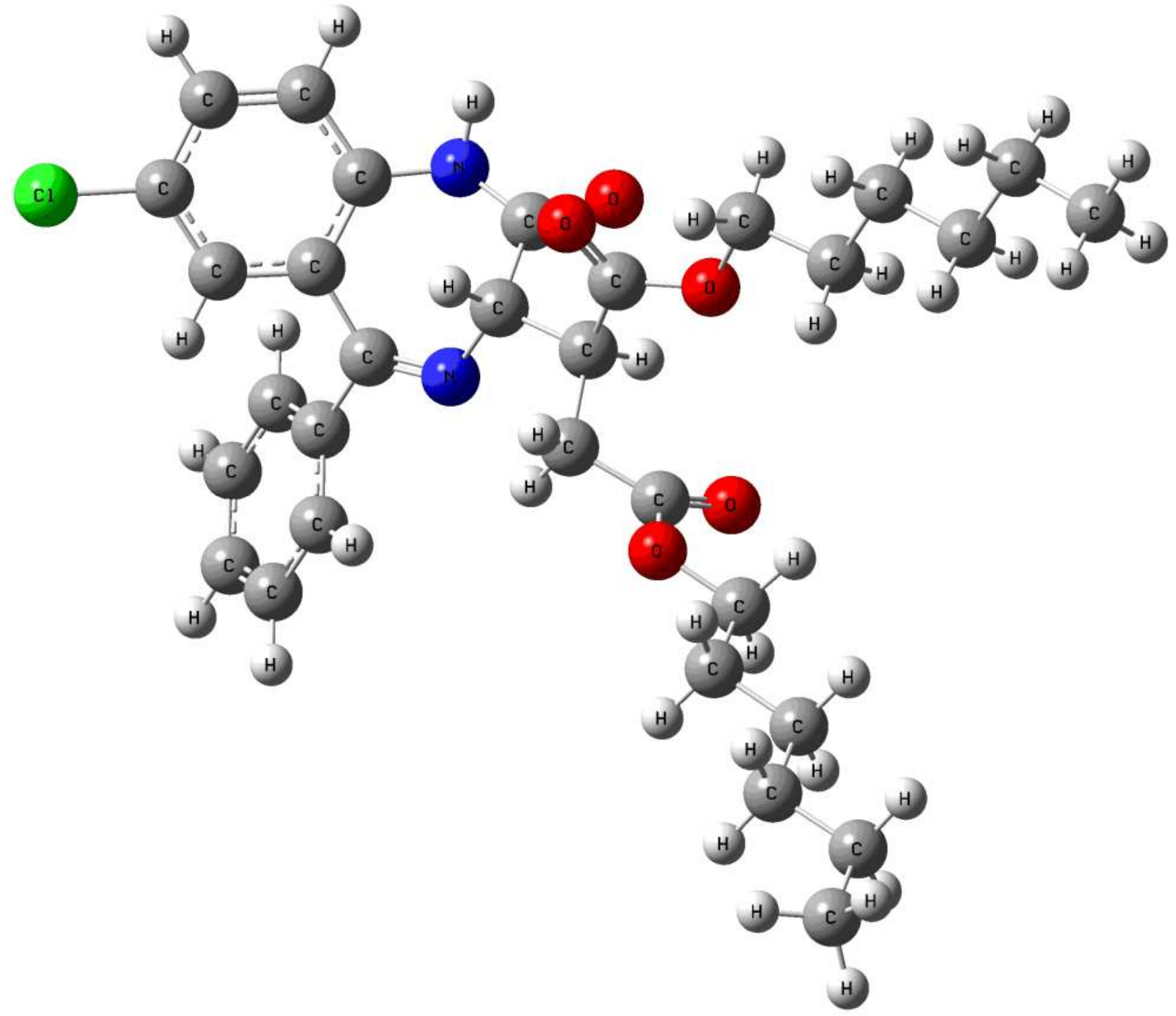

**Figure 4 : Optimied structure of C-alkylated product by $AM_1$.**

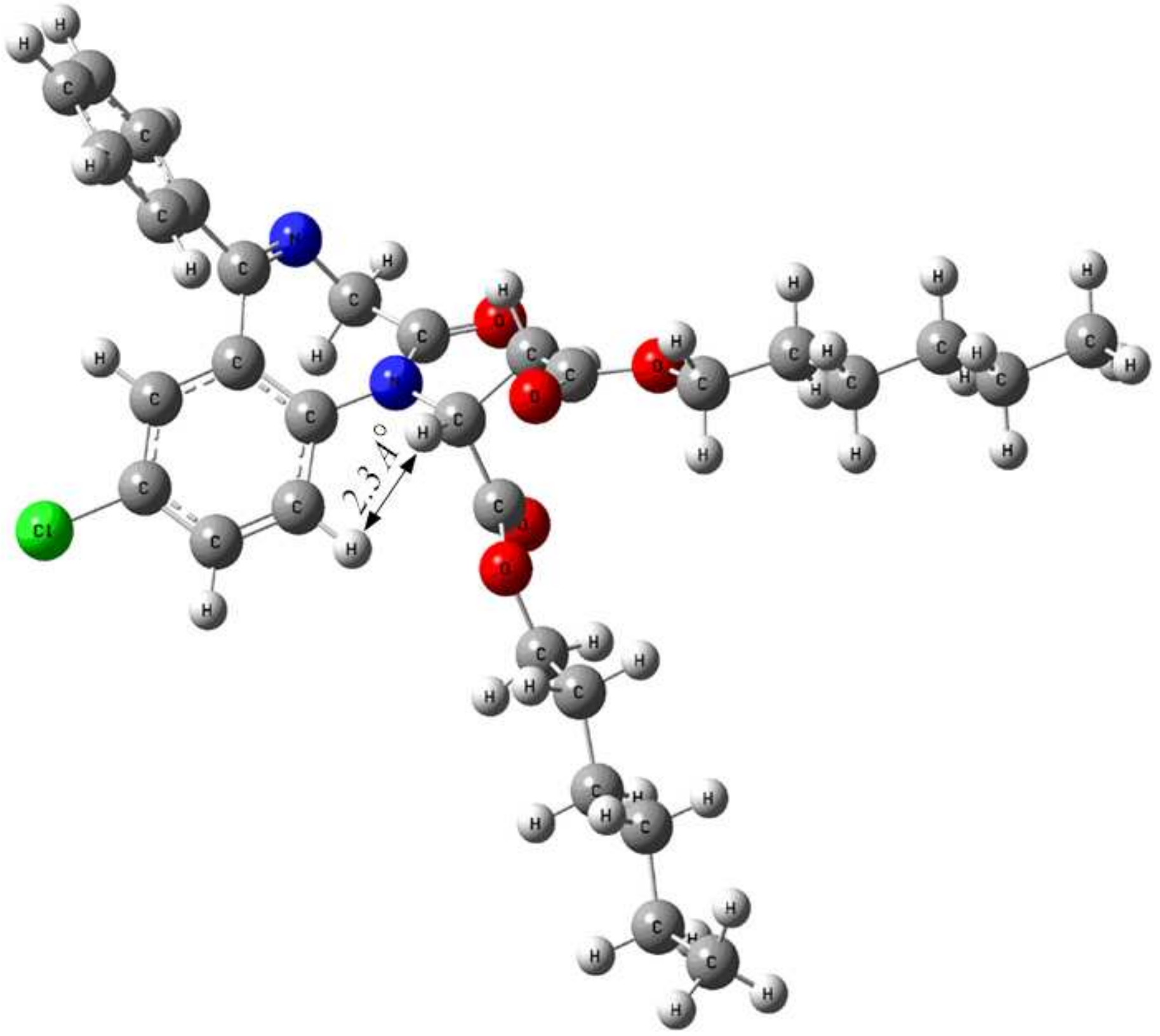

**Figure 5 : Optimied structure of N-alkylated product by $AM_1$.**

- **Study $^{1}$H-NMR spectrum of the 1,4-benzodiazepine-2-one addition reaction with iso-propyl fumarat**

  The Figure 6 shows $^{1}$H-NMR spectrum of the product of additional reaction of 7-chloro-1,3-dihydro-5-phenyl-2H-1,4-benzodiazepine-2-one molecule with iso-propyl fumarat.

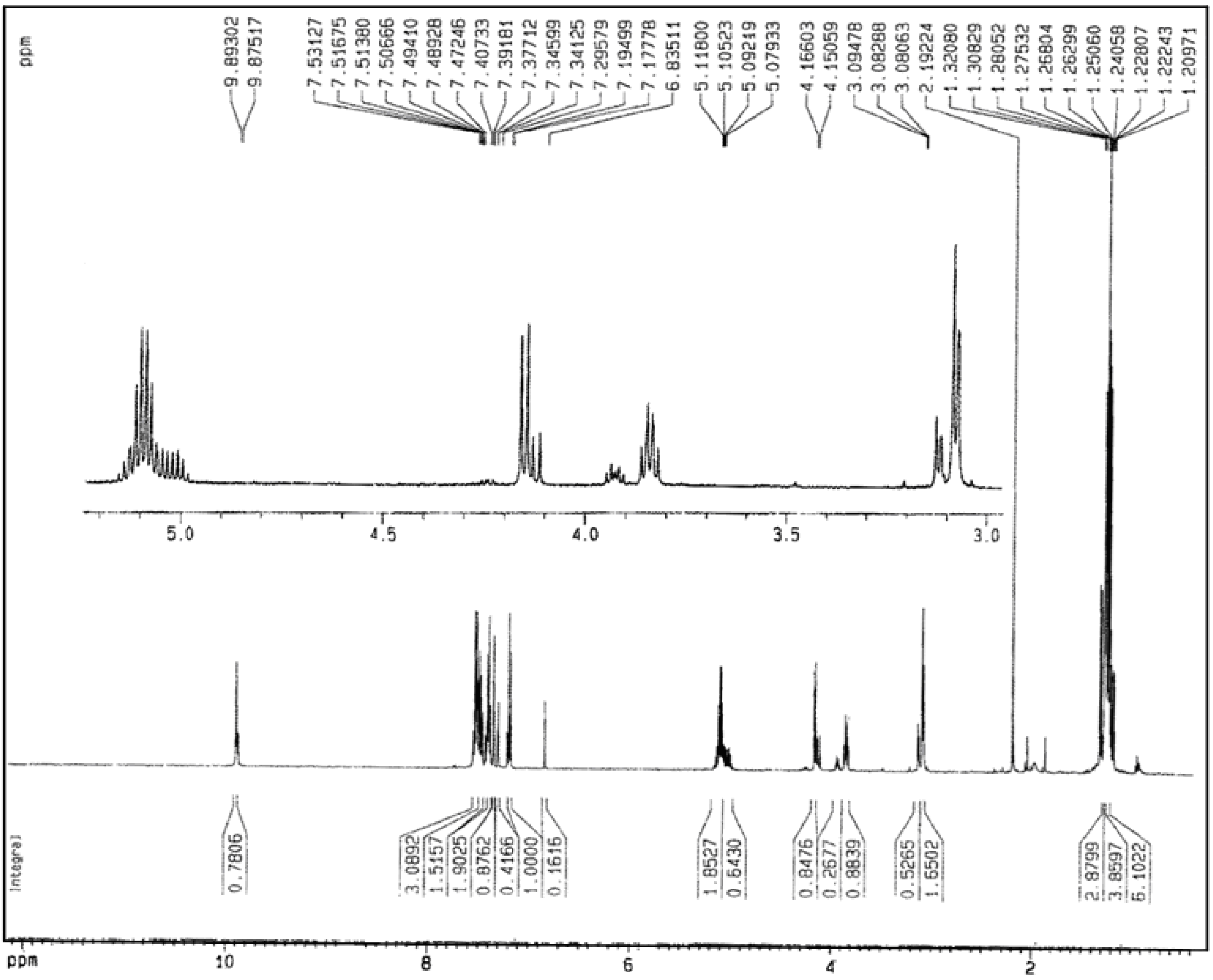

**Figure 6: $^{1}$H-NMR spectra of additional product of 7-chloro-1,3-dihydro-5-phenyl-2H-1, 4-benzodiazepine-2-one with iso-propyl fumarat.**

In this reaction is created two diasteromers based on this spectrum. This is confirmed by calculation of the peaks intensity ratio of each two adjacent signals that values of the ratios are equal together . The ratio of two adjacent doublet signals in region δ=3.08 ppm and δ=3.10 ppm is equal to 3.1. The ratio of two adjacent quartet signals in region δ=3.80 ppm and δ=3.90 ppm is equal to 3.3. The ratio of two adjacent multiplet signals in region δ=5.05 ppm and δ =5.10 ppm is equal to 2.88.

The above results indicate that one of the diasteromers is generated three times more than another diasteromer. Also, we obtain major diasteromer percent based on the peaks intensity. The major diasteromer percents are 75.80 %, 76.75% and 74.20%, respectively (see Figure 7).

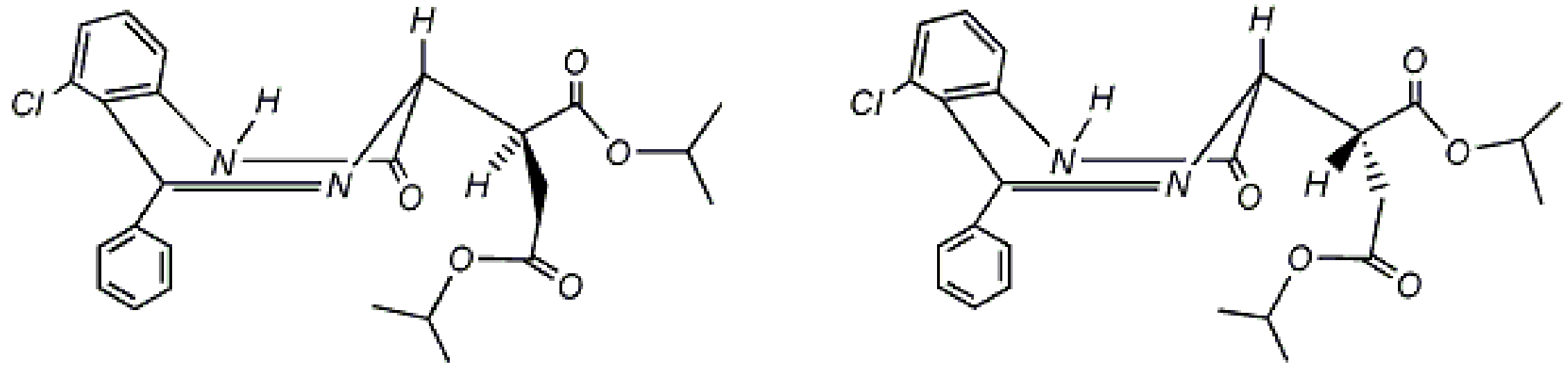

**Figure 7: Two diasteromers were produced from additional reaction of 7-chloro-1,3-dihydro-5-phenyl-2H-1,4-benzodiazepine-2-one with iso-propyl fumarat.**